\documentclass{jfm}

\usepackage{amsmath}
\usepackage{amsfonts}
\usepackage{amssymb}
\usepackage{enumitem}
\usepackage{mathtools}
\usepackage{natbib}
\usepackage{physics}
\usepackage{color}
\usepackage{upgreek}

\newcommand{\Dt}{{\rm D}_t}

\newcommand{\rmd}{{\rm d}}

\newcommand{\vg}[1] {{\boldsymbol{#1}}}

\DeclareMathOperator{\J1}{J_1}

\DeclareMathOperator{\JJ}{J}

\DeclareMathOperator{\PP}{P}
\DeclareMathOperator{\RR}{R}

\newcommand{\half}{\tfrac{1}{2}}

\title{Oceanic dipoles in a surface quasigeostrophic model}
\shorttitle{Surface-intensified mesoscale oceanic dipoles}
\author{E. R. Johnson \& M. N. Crowe}
\affiliation{Department of Mathematics, University College London, London, WC1E 6BT, UK}
\date{}

\begin{document}

\maketitle

\begin{abstract}
Analysis of satellite altimetry and Argo float data leads Ni {\em et al.} (2020, JGR Oceans, 10.1029/2020JC016479) to argue that mesoscale dipoles are widespread features of the global ocean having a relatively uniform three‐dimensional structure that can lead to strong vertical exchanges. Almost all the features of the composite dipole they construct can be derived from a model for multipoles in the surface quasigeostrophic equations for which we present a straightforward novel solution in terms of an explicit linear eigenvalue problem, allowing simple evaluation of the higher radial modes that appear to be present in the observations and suggesting that mass conservation may explain the observed frontogenetic velocities.
\end{abstract}

\section{Introduction}
\begin{figure}
\includegraphics[width=0.49\textwidth]{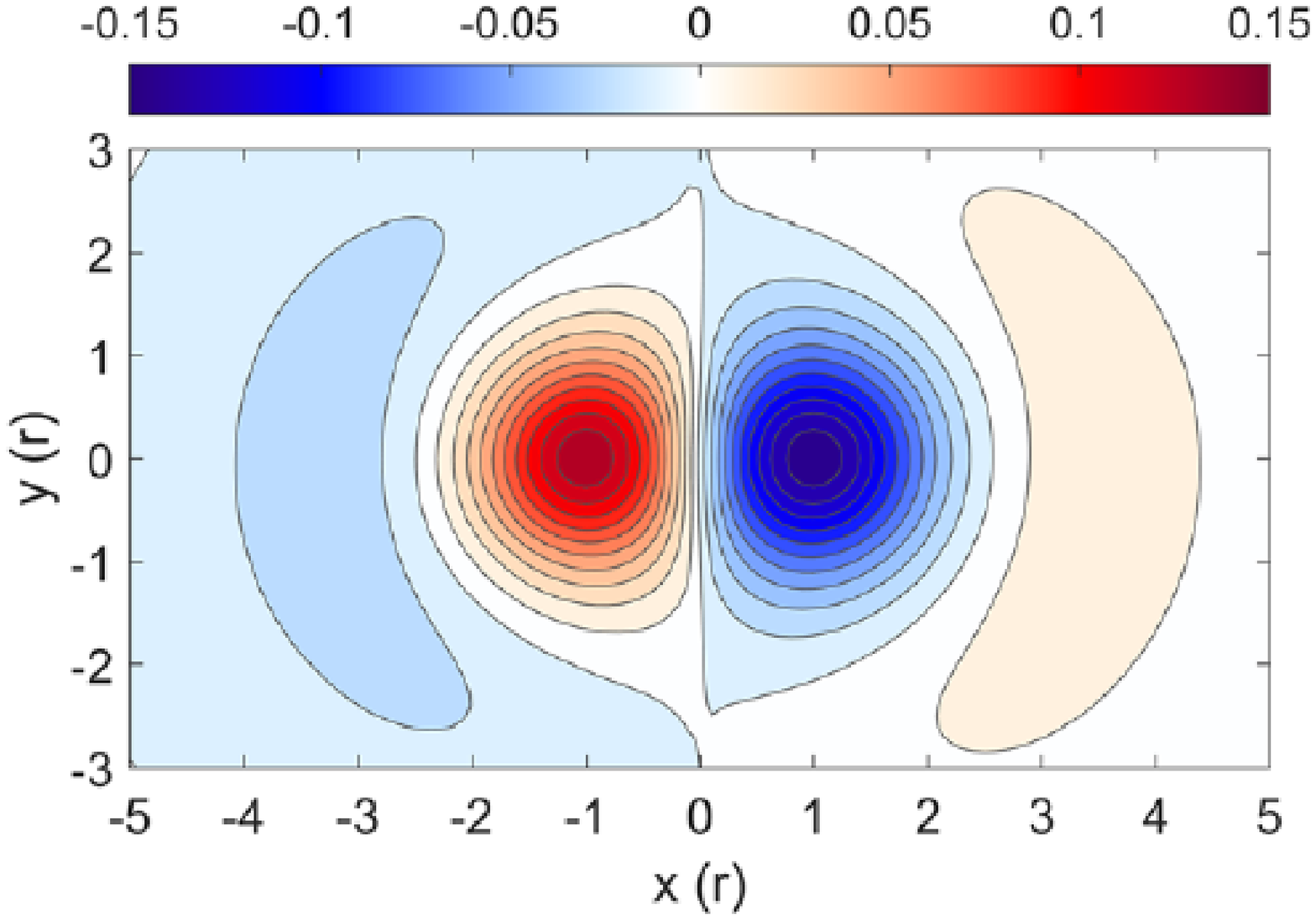}\hfill
\includegraphics[width=0.49\textwidth]{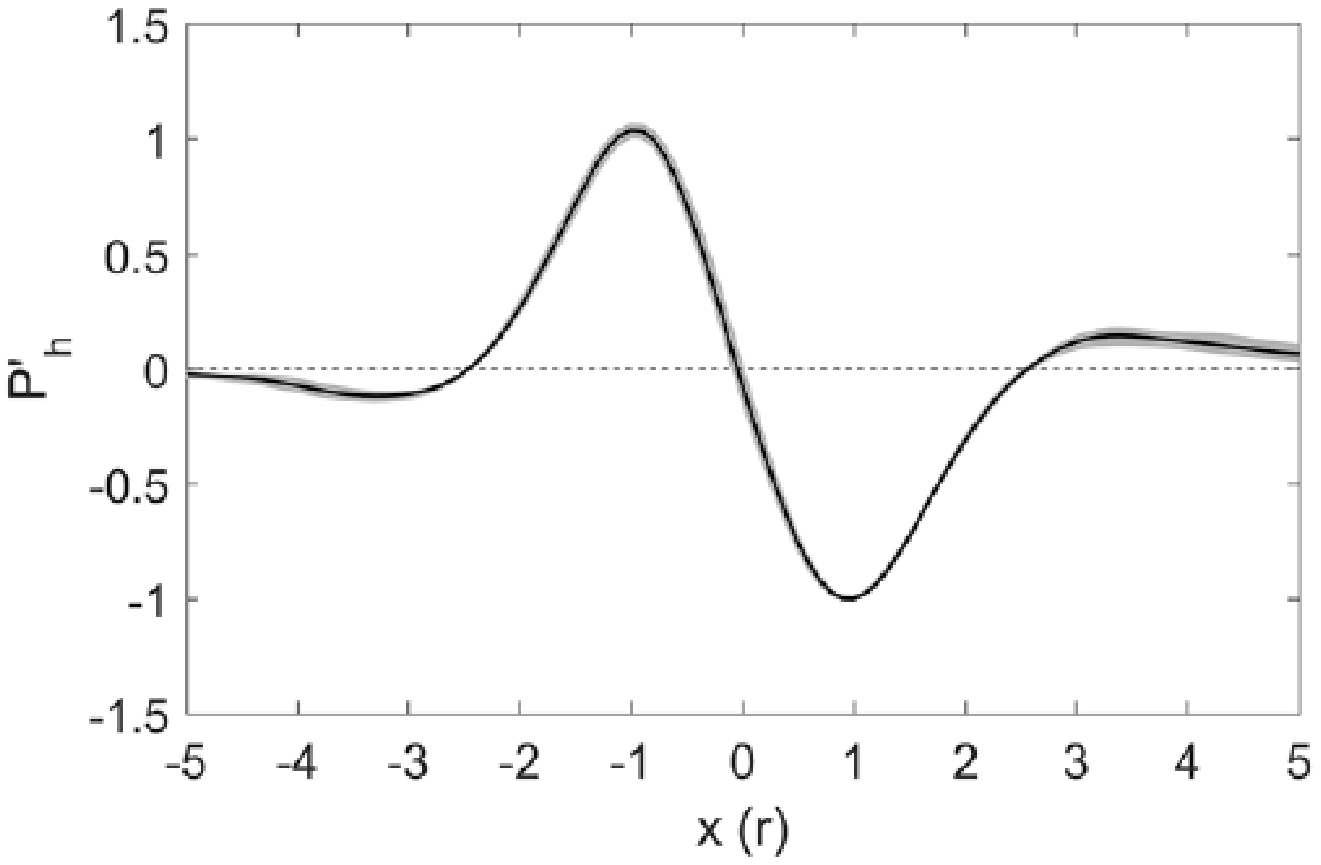}\\
\hspace*{0.25\textwidth}(a)\hspace{0.50\textwidth}(b)\\
\includegraphics[width=0.51\textwidth]{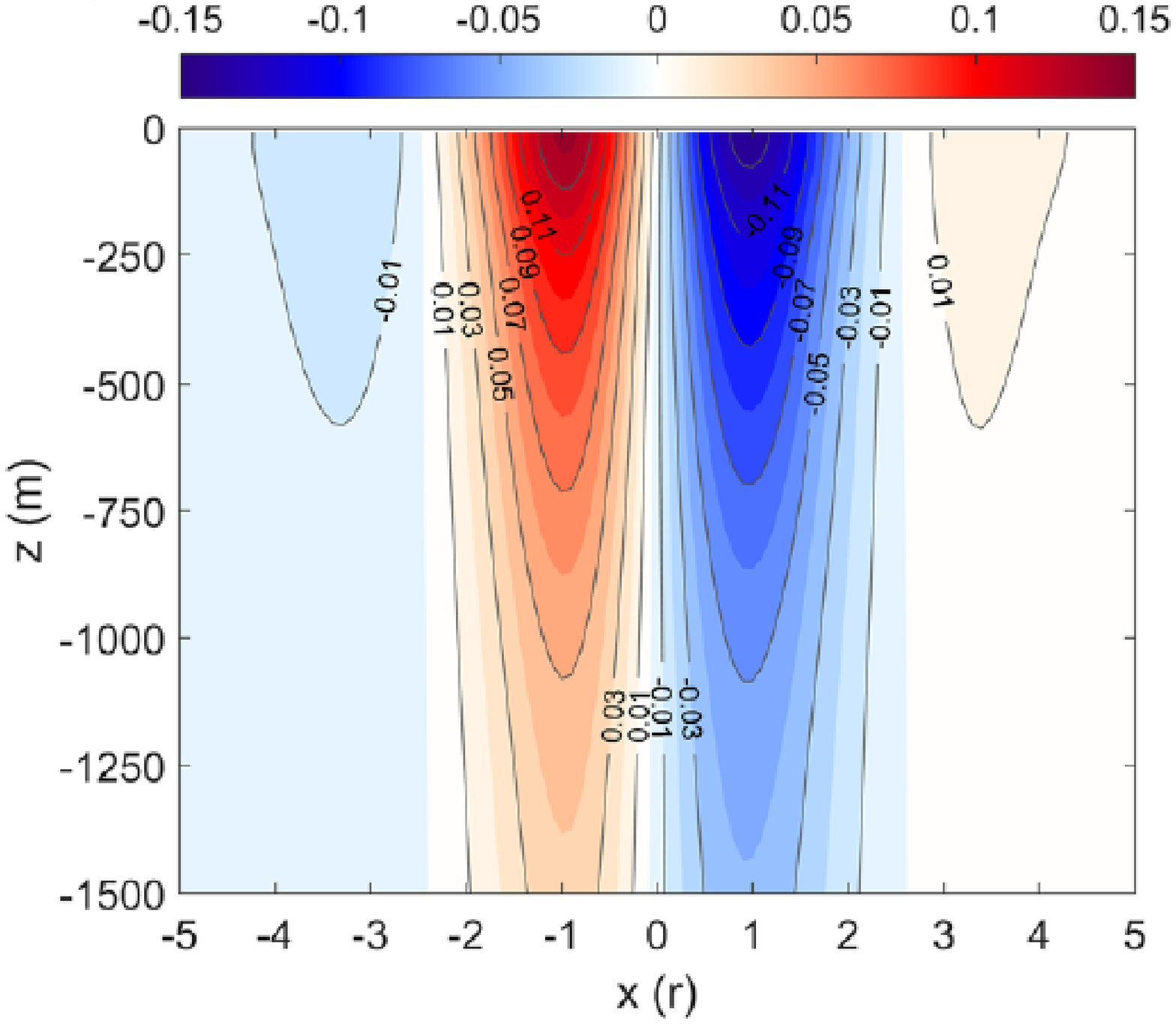}\hfill
\includegraphics[width=0.47\textwidth]{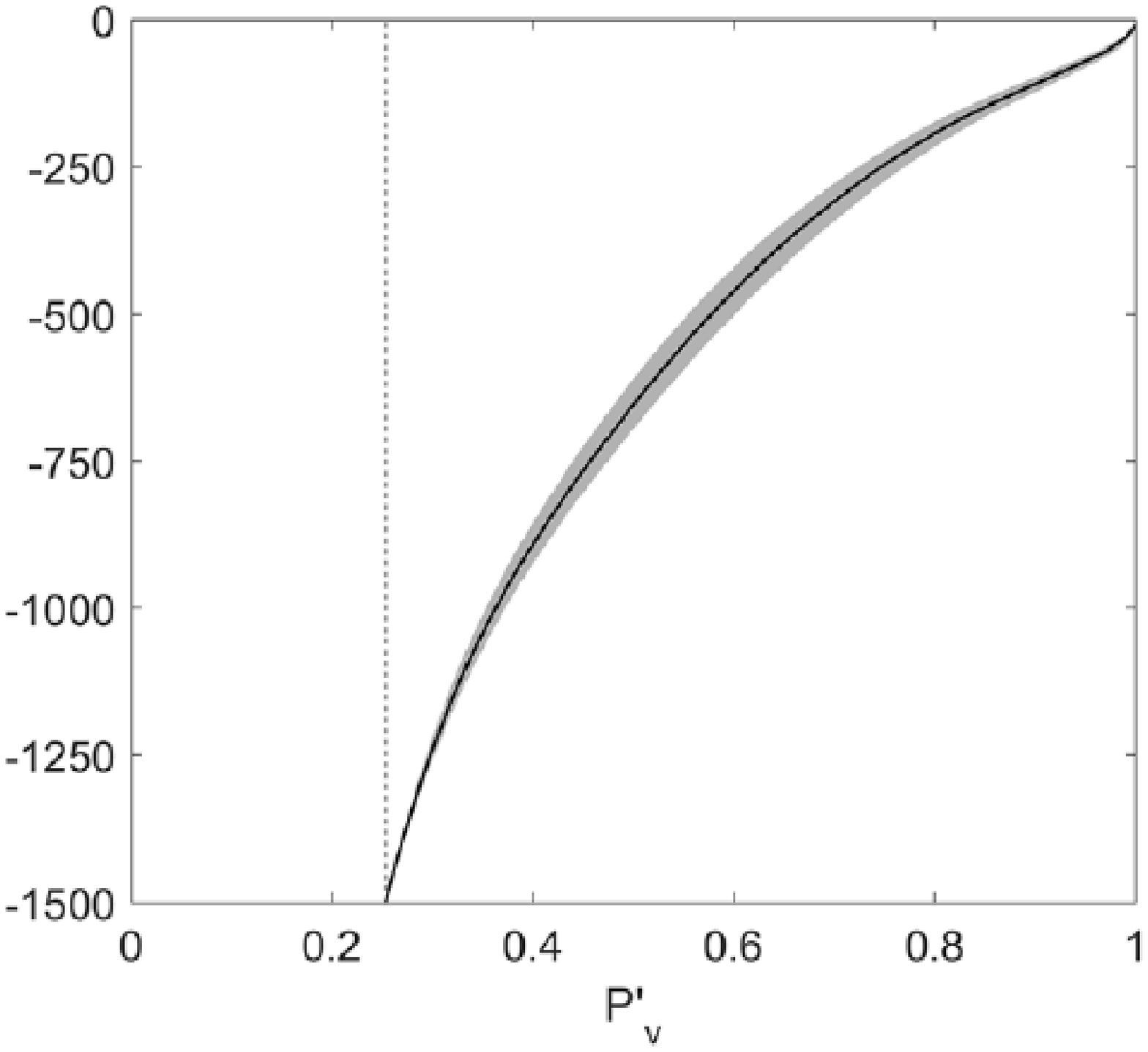} 
\hspace*{0.25\textwidth}(c)\hspace{0.50\textwidth}(d)\\
\caption{The composite dipole of  \cite{NiZWH20}. (a) Composite average of surface pressure anomalies in  dipole‐centric coordinates.  (b) Cross-sections along $y=0$ of the surface pressure normalized by its magnitude. (c) A vertical cross-section along $y=0$ of the composite dipole. (d) The vertical pressure variation (normalised
by its surface value) inside the composite dipole.
}
\label{f:Ni}
\end{figure}

\cite{NiZWH20} (NZWH herein) point out that mesoscale eddies account for the majority of the ocean's kinetic energy noting further that dipolar eddies are perhaps the simplest dynamically consistent and potentially ubiquitous features in the ocean. NZWH observe from satellite altimetry and Argo float data that these dipoles have a relatively uniform, surface-intensified, three‐dimensional structure and propose a composite dipole structure (NZWH, figure 5) reproduced in figure \ref{f:Ni} here. From their analyses they note that dipoles promote the vertical motions vital for supplying nutrients to the euphotic zone to support primary production and for sequestering carbon to the deep ocean.

The goal of the present study is to show that these observations can be closely modelled by dipole modes of the surface quasigeostrophic equations derived from a linear reformulation of a model presented by \cite{MurakiS07} (MS herein) and solved there by approximation of a transform function that decays slowly over an infinite interval, integration against a slowly decaying kernel function and then a nonlinear iterative search for a root. The model differs from that of \cite{LahayeZD20} who consider dipoles in the thermal quasi-geostrophic equations, where, as they note, fluid moves as vertical columns and so the motion is either barotropic, stretching from ocean surface to floor, or confined to the upper layer of a $1\tfrac{1}{2}$-layer flow \cite[][]{WarnefordD13}, making extraction of vertical velocity fields more difficult. Section 2 summarises the MS derivation of the governing dual integral equations and then demonstrates how the problem can be expressed as the solution of a linear eigenvalue problem with simple explicit coefficients.  Numerical integrations of the nonlinear time-dependent governing equations show that the lowest mode dipole is long-lived and the mode-two dipole breaks up after propagating a distance of order three or four radii. Section 3 argues that the composite dipole of NZWH might be a combination of mode-one and mode-two observations and shows that the observations are accurately reproduced by an 80/20 mode-one/mode-two composite model dipole. In particular, the horizontal and vertical structures of the vertical velocity field of the composite model dipole reproduce the reported observations. 

\section{The governing equations and dipole solutions}\label{S:gov}

The observed dipoles are of limited latitudinal extent and motions are velocities are sufficiently small that the Rossby number $\epsilon=U/fL$ of the flow, for $U$ the dipole propagation speed, $L$ a typical eddy width and $f$ the constant Coriolis parameter, can be taken to be small. Take the ocean to be of depth $H$ and let the stably stratified background density profile have buoyancy frequency $N_0 N(z)$ where $N(z)$ has maximum unity.  In Cartesian coordinates $Oxyz$ with velocity components $(u,v,w)$ the leading order (in $\epsilon$) nondimensional inviscid momentum equations are the geostrophic and hydrostatic equations
\begin{equation}
    u=-p_y, \qquad v=p_x, \qquad \sigma=p_z,
\end{equation}
with $(x,y)$ are scaled on $L$, $z$ on $fL/N_0$, $(u,v)$ on $U$, $p$ on $\rho_0fUL$ for some representative density $\rho_0$ and $\sigma$ is the buoyancy acceleration scaled on $UN_0$. The equation for the conservation of density in the absence of diffusion is
\begin{equation}\label{density}
   \Dt\sigma + N^2w = 0, 
\end{equation}
with $\Dt=\partial_t + u\partial_x+v\partial_y$, for time $t$ scaled on $L/U$, and this balance between horizontal advection of disturbance density and vertical advection of background density giving the scale of $w$ as $U^2/N_0L=\epsilon U f/N_0$. The system is closed by the continuity equation for incompressible flow, giving to leading order  \cite[][]{Johnson78b}
\begin{subequations}\label{gov1}
\begin{align}
\Dt[p_{xx}+p_{yy} + (N^{-2}p_z)_z] &=0, \label{gov_main}\\
\Dt[p_z]&=0, \qquad z=0, -B, \label{gov_surf}
\end{align}
\end{subequations}
where $B=N_0H/fL$. The boundary conditions \eqref{gov_surf} come from \eqref{density} and requiring that $w=0$ on the top and bottom boundaries with the vanishing of $w$ on the upper boundary following by noting that the horizontal extent of the motion is sufficiently small compared to the external Rossby radius of deformation, $(gH)^{1/2}/f (\approx 2000$km in the open ocean) that the ocean surface can be taken as rigid. Equation \eqref{gov_main} describes the advection of interior potential vorticity (iPV) and \eqref{gov_surf} states that surface potential vorticity (sPV), and thus buoyancy acceleration, is advected along the upper and lower boundaries. The pressure field, and thus the velocity, are obtained at each instant from the iPV and the sPV through an appropriate Dirichlet-Neumann operator. Various forms of this operator are given in \cite{Johnson78b} where they are used to construct steady nonlinear eddies in the neighbourhood of seamounts. The temporal evolution of these surface geostrophic equations has subsequently been discussed by \cite{HeldPGS95}. 

The dipoles described by NZWH decay monotonically away from a thin surface layer and appear to be uninfluenced by the ocean floor. It is thus sufficient initially to consider infinitely deep flows ($B\to\infty$) where the iPV vanishes, as it does for all flows started from rest, with the maximum principle enforcing the monotonic interior decay. The horizontal orientation of the dipole in NZWH is arbitrary and so for ease of comparison with MS we rotate the axes by $\pi/2$ anticlockwise and consider a dipole advancing steadily in the positive-$x$ direction at constant scaled speed unity into a region of undisturbed fluid. Then in a frame moving with the dipole the flow is steady with uniform velocity  $-\hat{\vg{x}}$
at large distances. System \eqref{gov1} then becomes, for zero iPV, 
\begin{subequations}\label{gov2}
\begin{align}
p_{xx}+p_{yy} + (N^{-2}p_z)_z &=0, \label{gov2_main}\\
\partial(y+p,p_z) &=0, \qquad &z=0, \label{gov2_surf}\\
p_z &\to 0, &z\to -\infty, \label{gov2_deep}
\end{align}
\end{subequations}
where $\partial(.,.)$ denotes the Jacobian. 

Simple fully nonlinear solutions of \eqref{gov2} follow by considering solutions where the sPV vanishes outside the modon and looking for a Lamb-Chaplygin-like \cite[][]{MeleshkoH94} solution where the sPV inside the modon is a linear function of the total streamfunction $y+p$. The observed dipoles appear to show an exponential decay with depth and so attention is restricted to uniform stratification with $N(z)\equiv1$\footnote{The form of the solution for finite depth and non-uniform stratification is noted briefly in Appendix \ref{S:expstrat}}. Consider circular dipoles of radius unity so
\begin{equation}\label{sbc}
    p_z=
    \begin{cases}
     K(p + y) \qquad &x^2+y^2<1 \\
    0,  &x^2+y^2>1,
    \end{cases}
\end{equation}
where $K$ is a constant wavenumber, determined as part of the solution. For steady solutions the boundary of the non-zero sPV is a streamline,
\begin{equation}\label{bdry}
   p + y = 0 \qquad\text{on } x^2+y^2=1,\; z=0.
\end{equation}
To solve the system of \eqref{gov2_main}, \eqref{sbc} and \eqref{bdry}, follow MS, introducing polar coordinates $(x,y)=(r\cos\phi,r\sin\phi)$ and look for solutions of the form
\begin{equation} \label{hankel}
 p(r,\phi,z) = \sin\phi\int_0^\infty \hat{p}(\xi)\J1(\xi r)\exp(\xi z)\xi \;\rmd\xi,
\end{equation}
so $\hat{p}$ is the Hankel transform of $p(r,\pi/2,0)$. 
This form satisfies the governing interior equation. MS obtain an integral equation for $\hat{p}$ and by discretising at 512 points obtain an algebraically decaying representation of $\hat{p}$, evaluating the infinite integrals numerically and using a nonlinear root-finding method to obtain $K$. The method here gives an economical series solution representation of the solution, with the coefficients and the wavenumber $K$ determined simultaneously as the solution of an explicit standard linear eigenvalue problem with known rational coefficients by following Tranter's method \cite[][p.111]{Tranter71} as in \cite{HockingMW79}.  The method below achieves the same accuracy as MS using only 12 terms with the smallest neglected coefficient of order 10$^{-6}$.  

Substituting \eqref{hankel} in \eqref{sbc} gives the dual integral equations 
\begin{subequations}
\begin{align}
\int_0^\infty A(\xi)\J1(\xi r)\;\rmd\xi-K\int_0^\infty \xi^{-1}A(\xi)\J1(\xi r)\;\rmd\xi &= K r, &0\leq r\leq1,\label{sbc2} \\
\int_0^\infty A(\xi)\J1(\xi r)\;\rmd\xi &= 0, &r\geq1, \label{outer}
\end{align}
\end{subequations}
where $A(\xi)=\xi^2\hat{p}(\xi)$. Tranter's method consists of looking for a solution for $A(\xi)$ as a sum of terms of the form $\xi^{1-k}\JJ_{2n+1+k}(\xi)$, since terms of this form satisfy \eqref{outer} identically. Choosing $k=1$ ensures that $p_z$ vanishes at $r=0$ and is bounded at $r=1^-$. Thus look for a solution of the form
\begin{equation}\label{series}
 A(\xi) = \sum_{n=0}^\infty a_n \JJ_{2n+2}(\xi),
\end{equation}
where the coefficients $a_n$ are  to be determined. Now 
\begin{equation}\label{int1}
 \int_0^\infty \JJ_{2n+2}(\xi)\J1(\xi r) \;\rmd\xi =
 \begin{cases}
  \RR_n(r), &r<1\\
  0 &r>1,
 \end{cases}
\end{equation}
where $\RR_n(r)$ is the Zernike radial function $(-1)^n\mathcal{R}^1_{2n+1}(r)$ from the diffraction theory of aberration \cite[][ch. 9]{BornW19}, a polynomial of degree $2n+1$, given by the the shifted Jacobi polynomial, 
\begin{equation}
 \mathcal{R}^1_{2n+1}(r)= r\PP^{(0,1)}_n(2r^2-1) = \sum_{k=0}^n\frac{(-1)^k(2n+1-k)!}{k!(n+1-k)!(n-k)!}r^{2n+1-k}.
\end{equation}
Equation \eqref{outer} is satisfied identically as expected and the problem reduces to obtaining the eigenvalue $K$ and coefficients $a_n$ so that \eqref{series} satisfies \eqref{sbc2}.  Substituting \eqref{series} in \eqref{sbc2} gives
\begin{equation}
\sum_{n=0}^\infty a_n [\int_0^\infty \JJ_{2n+2}(\xi) \J1(\xi r)\;\rmd\xi-K\int_0^\infty \xi^{-1}\JJ_{2n+2}(\xi) \J1(\xi r)\;\rmd\xi] = K r, \qquad 0\leq r\leq1.
\label{subs}
\end{equation}
Equation \eqref{int1} provides the Hankel transform of $\JJ_{2n+2}(\xi)/\xi$ so that inverting \eqref{int1}     
gives
\begin{equation}
 \JJ_{2n+2}(\xi)/\xi = \int_0^1 r\RR_{n}(r)\J1(\xi r)\;\rmd r.
\end{equation}
The functions $\RR_n$ are orthogonal over $0\leq r\leq 1$ with weight $r$ and so multiplying \eqref{subs} by $4(m+1)r\RR_{m}(r)$ and integrating from 0 to 1 gives the linear equation
\begin{equation}
 ({\bf B} - K {\bf C}) \vg{a} = K\vg{c},
 \label{inter}
 \end{equation}
where {\bf B} and {\bf C} are the matrices and $\vg{c}$ the vector with components
\begin{align}
 b_{mn} &=4(m+1)\int_0^\infty \xi^{-1}\JJ_{2m+2}(\xi)\JJ_{2n+2}(\xi)\;\rmd\xi = \delta_{mn},  \label{bmndef} \\
 c_{mn} &= 4(m+1)\int_0^\infty \xi^{-2}\JJ_{2m+2}(\xi)\JJ_{2n+2}(\xi)\;\rmd\xi \\
        &=\frac{16(m+1)(-1)^{m-n+1}}{[(2m-2n-1)(2m-2n+1)(2m+2n+3)(2m+2n+5)\pi]}, \\
 c_m &=4(m+1)\int_0^1 r^2\RR_{m}(r)\;\rmd r = \delta_{m0}, \label{cmdef}
\end{align}
where $\delta_{mn}$ is the Kronecker delta. Thus \eqref{inter} becomes 
\begin{equation}
 [{\bf C} - K^{-1}{\bf I}] \vg{a} = -\vg{c},
\label{inter2}
 \end{equation}
where {\bf I} is the identity matrix. In general, solving inhomogeneous eigenvalue problems like \eqref{inter2} is not straightforward. However the inhomogeneity is confined here to the first row of \eqref{inter2} and this can be treated separately. First note that in general the solution given by \eqref{series} gives a discontinuity in $p_z$ across $r=1$.
In order that the solution is continuous there
\begin{equation}
 \sum_{n=0}^\infty a_n \int_0^\infty \JJ_{2n+2}(\xi)\J1(\xi r) \;\rmd\xi = \sum_{n=0}^\infty (-1)^n a_n = 0, \quad\text{i.e. }\quad a_0 = \sum_{n=1}^\infty (-1)^{n-1} a_n.
 \label{a0}
\end{equation}
Hence $a_0$ can be replaced in rows below the first in \eqref{inter2} to give a linear homogeneous eigenvalue problem for $\vg{\hat{a}} =\{a_n\}_{n\geq1}$, with coefficient matrix with elements
\begin{equation}
     \hat{c}_{mn}= c_{mn} +(-1)^nc_{m0}, \qquad (m\geq1, n\geq1).
\end{equation}
Truncating the sum \eqref{series} at $n=N=12$ and solving by any standard method gives a smallest eigenvalue of $K=4.1213$, as obtained by nonlinear root-finding in MS. Higher eigenvalues correspond to modon solutions with interior circular nodal lines. The vector $\vg{a}$ is thus determined to within a multiplicative constant whose value follows from satisfying the first row of \eqref{inter2},
\begin{equation}
 (c_{00}-K^{-1})a_0 + \sum_{n=1}^\infty c_{0n} a_n = 1.
\end{equation}
This completes the solution, giving the surface buoyancy as a sum of Zernike polynomials in $x$ and $y$,
\begin{equation}\label{pzZ}
     p_z(x,y,0) = 
    \begin{cases}
    \sin\phi\sum_{n=0}^\infty a_n \RR_n(r) \qquad &r\leq a\\
    0 & r\geq a.
    \end{cases}   
\end{equation}
Appendix \ref{S:series} shows that the sum \eqref{pzZ} can be evaluated directly from a simple three-term recurrence relation without computation of the Zernike polynomials.

The surface pressure can be expressed similarly as a sum of hypergeometric functions but for computational purposes it more straightforward to obtain $p(x,y,0)$ from \eqref{pzZ} by Fourier transforming in $x$ and $y$, to obtain $\hat{p_z}(k,l)$, for horizontal wavenumber $(k,l)$, obtaining the transform of $p(x,y,0)$ as
\begin{equation}
\hat{p} = \hat{p}_z/\sqrt{k^2+l^2},
\label{DNop}
\end{equation}
and inverting for $p$.
\begin{figure}
    \hfill
    \includegraphics[width=0.40\textwidth]{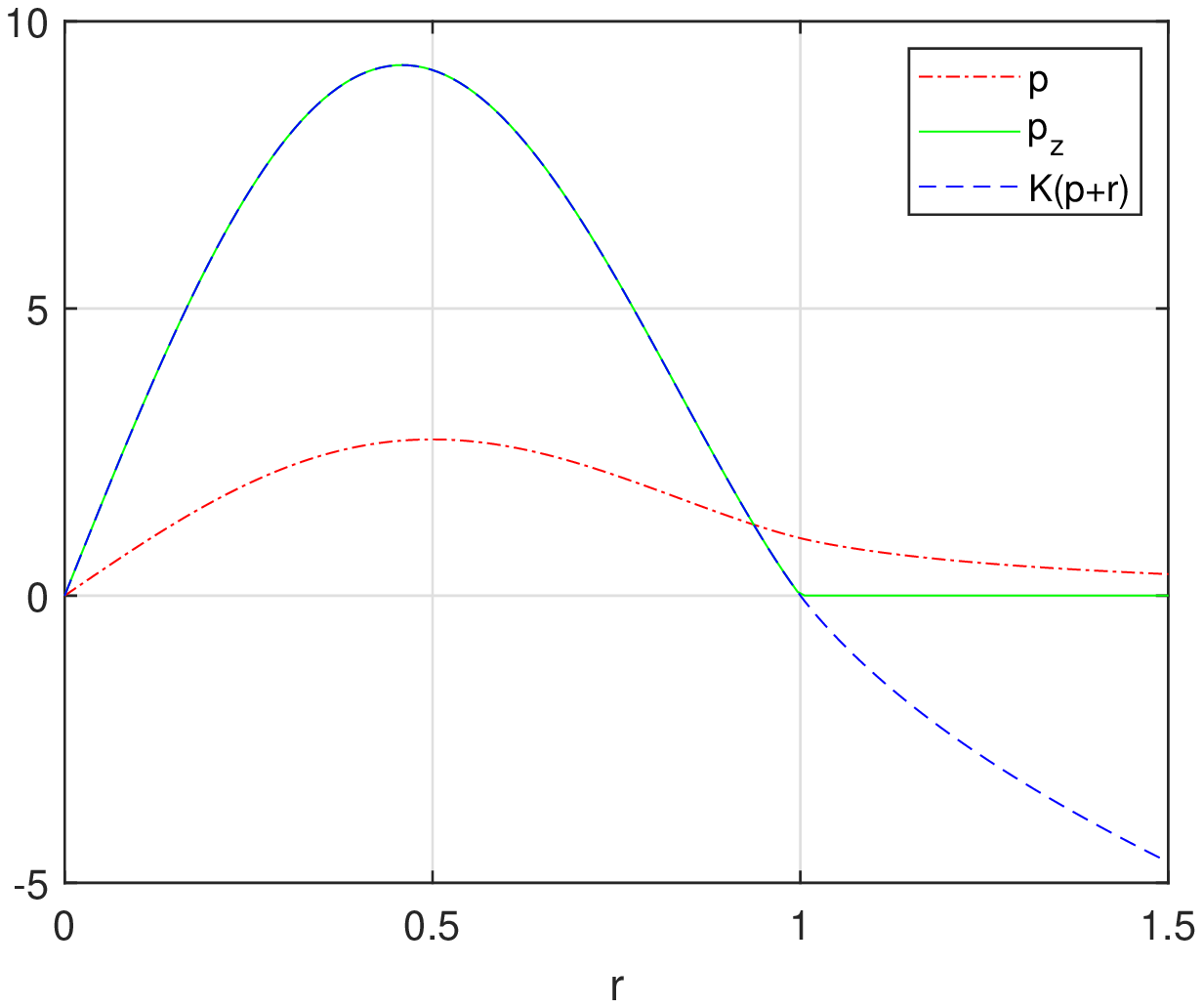}
    \hfill
    \includegraphics[width=0.40\textwidth]{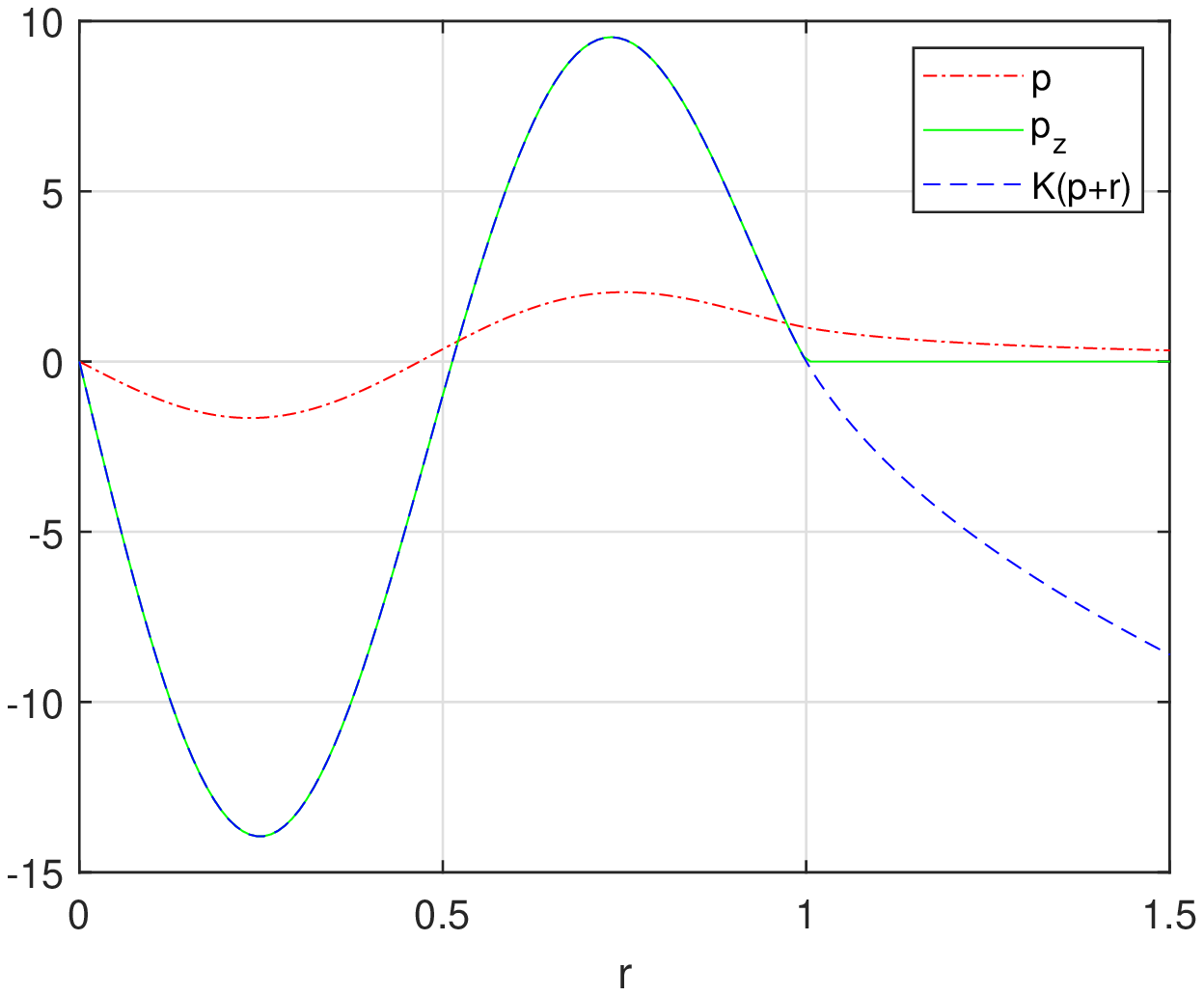}
    \hfill\phantom{.}\\
    \hspace*{0.25\textwidth}(a)\hspace{0.45\textwidth}(b)\\
    \hfill
    \includegraphics[width=0.48\textwidth]{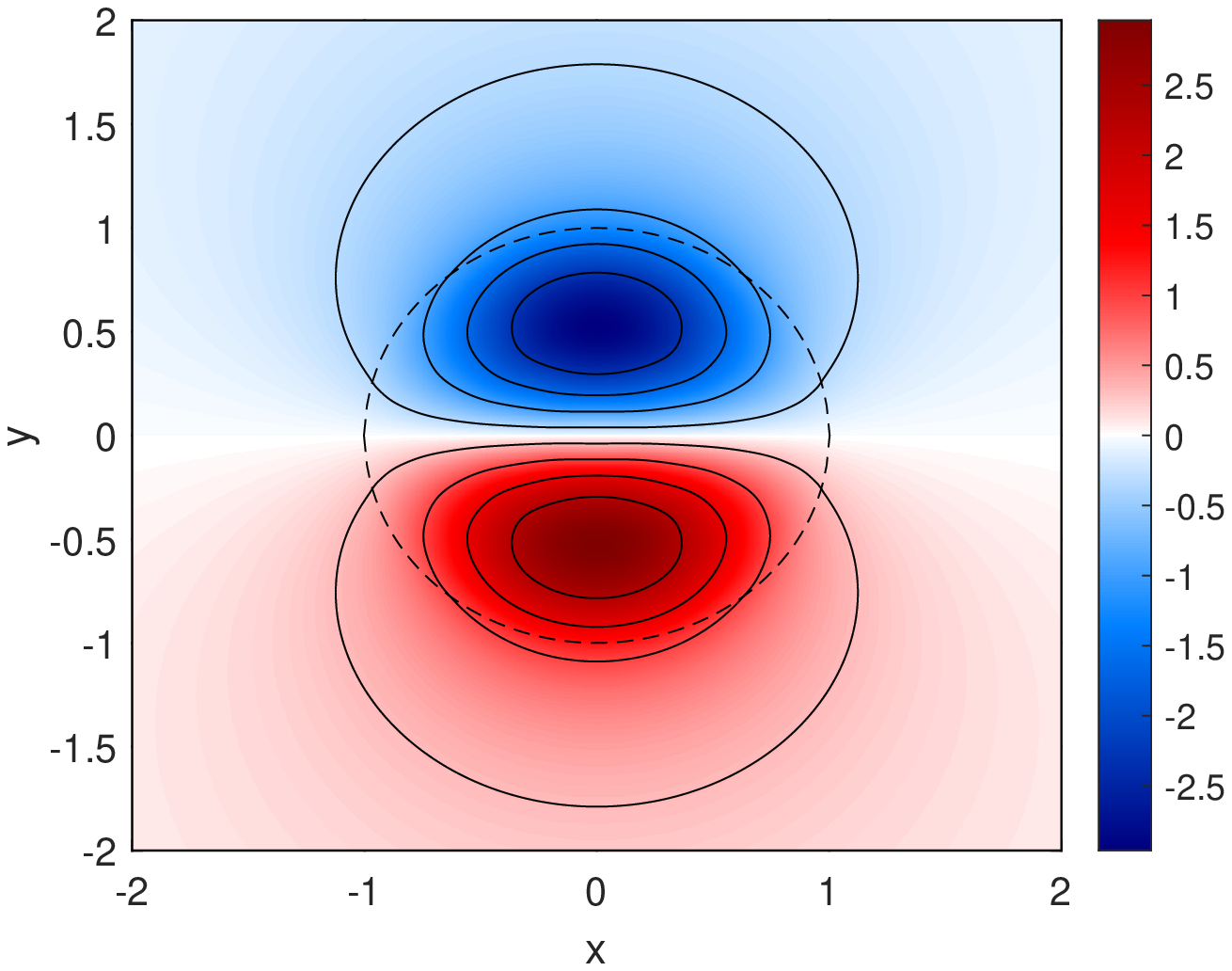}
    \hfill
    \includegraphics[width=0.48\textwidth]{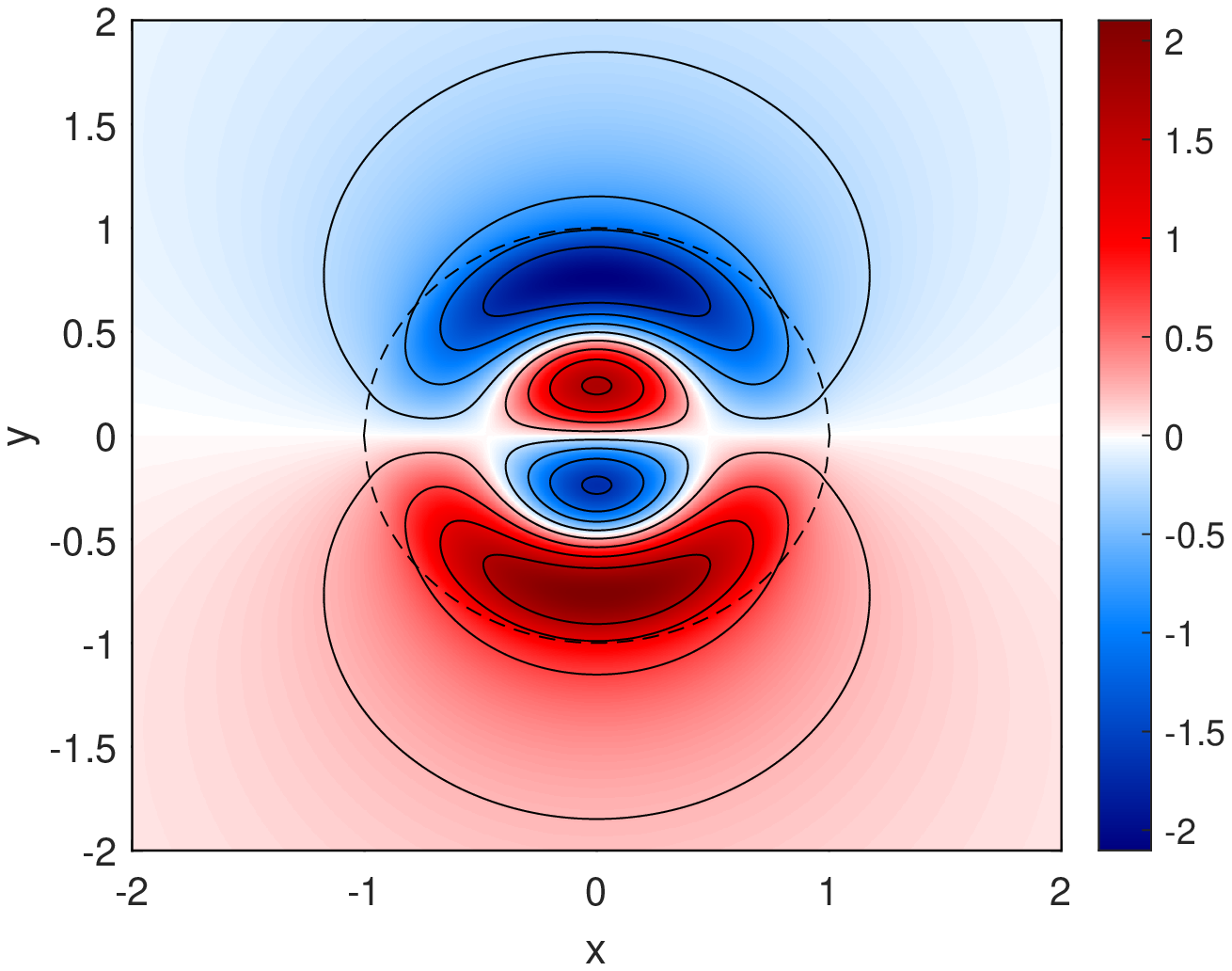}
    \hfill\phantom{.}\\
    \hspace*{0.25\textwidth}(c)\hspace{0.40\textwidth}(d)\\
\caption{$p$, $p_z$ and $K(p+r)$ along $x=0$ for (a) mode 1 (b) mode 2. (c,d) the corresponding surface pressure fields.
\label{f:m1}}
\end{figure}
Figure \ref{f:m1} shows $p$, $p_z$ and $K(p+r)$  and the corresponding surface pressure field for the lowest and second modes. The graphs of $p_z$ and $K(p+r)$ coincide for $r\leq1$ verifying that the solution indeed satisfies \eqref{sbc}.

The stability of these steadily propagating solutions can be investigated numerically using the Dedalus package \cite[][]{BurnsVOLB20} to solve system \eqref{gov1} for infinite depth and zero iPV, i.e. the unsteady version of system \eqref{gov2}, on the plane $z=0$ with $p$ obtained from $p_z$ at any instant through the Dirichlet-Neumann operator \eqref{DNop}, and initial conditions for $p$ and $p_z$ set by \eqref{pzZ}. We take a doubly periodic grid with $(x,y) \in [-25.6,25.6)\times[-25.6,25.6)$ using 2048 grid points in each direction. The domain size is chosen so that there is no significant influence on the vortex from the periodicity. Solutions are integrated for times $0\leq t\leq50$ using a $(3-\epsilon)$-order 3-stage Runge-Kutta scheme. A small hyperdiffusion term with hyperdiffusivity of $3.9\times10^{-10}$ is included for numerical stability.
\begin{figure}
     \includegraphics[width=\textwidth]{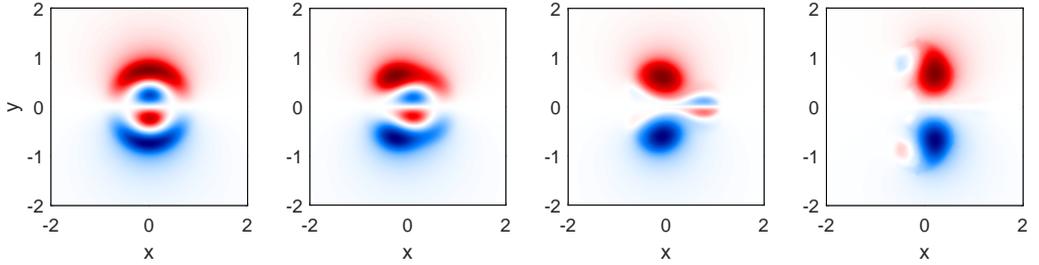}
\caption{The evolution of the surface buoyancy, $p_z$ of a mode-two modon at times (left to right) $t=2$, 3, 3.5 and 4. 
\label{f:mod2evol}}
\end{figure}
Long time integrations for mode one show no tendency to break up. The evolution for a mode-two vortex in figure \ref{f:mod2evol} shows that the vortex is unstable and breaks down in approximately $3.5$ time units, corresponding to the time taken to travel $3.5$ radii. This breakdown time varies weakly with grid resolution and hyperdiffusivity and solutions may break down faster in a full 3D simulation. The mode-two solutions here are unstable exact solutions of the equations of motion and so survive longer than the `coated' dipole solutions of \cite{Couder86} which do not satisfy the steady equations and immediately evolve away from their initial state. It is thus  possible that mode-two solutions may survive sufficiently long to be observed transiently in the ocean.  

Multiplying the governing equation by $y$ and integrating over the domain shows that the fluid impulse is conserved during the motion.
\begin{equation}
 \mu = -\int y p_z = -\pi \int_0^1 \sum_{n=0}^\infty a_n \RR_n(r)r^2\;\rmd r = -\tfrac{1}{4}\pi a_0.
\end{equation}
For mode 1 this gives $\mu\approx4.8744$ and for mode 2, $\mu\approx4.0257$. Similarly, multiplying the governing equation by $p$ and integrating over the domain shows that the fluid energy is conserved during the motion.
\begin{align}
 E &= \half\int p p_z = (\pi/2K) \int p_z^2 r \rmd r +\half\mu
 =  (\pi /2K) \int \hat{p_z}^2 \xi \rmd \xi  +\half\mu \\
 &= (\pi /8K)  \sum_{n=0}^\infty a_n^2/(n+1) + \half\mu,
 \end{align}
by \eqref{bmndef}. For mode 1 this gives $E\approx9.7488$ and for mode 2, $E\approx8.0513$.

\section{Modelling the observations}

\begin{figure}
\hfill
     \includegraphics[width=0.4\textwidth]{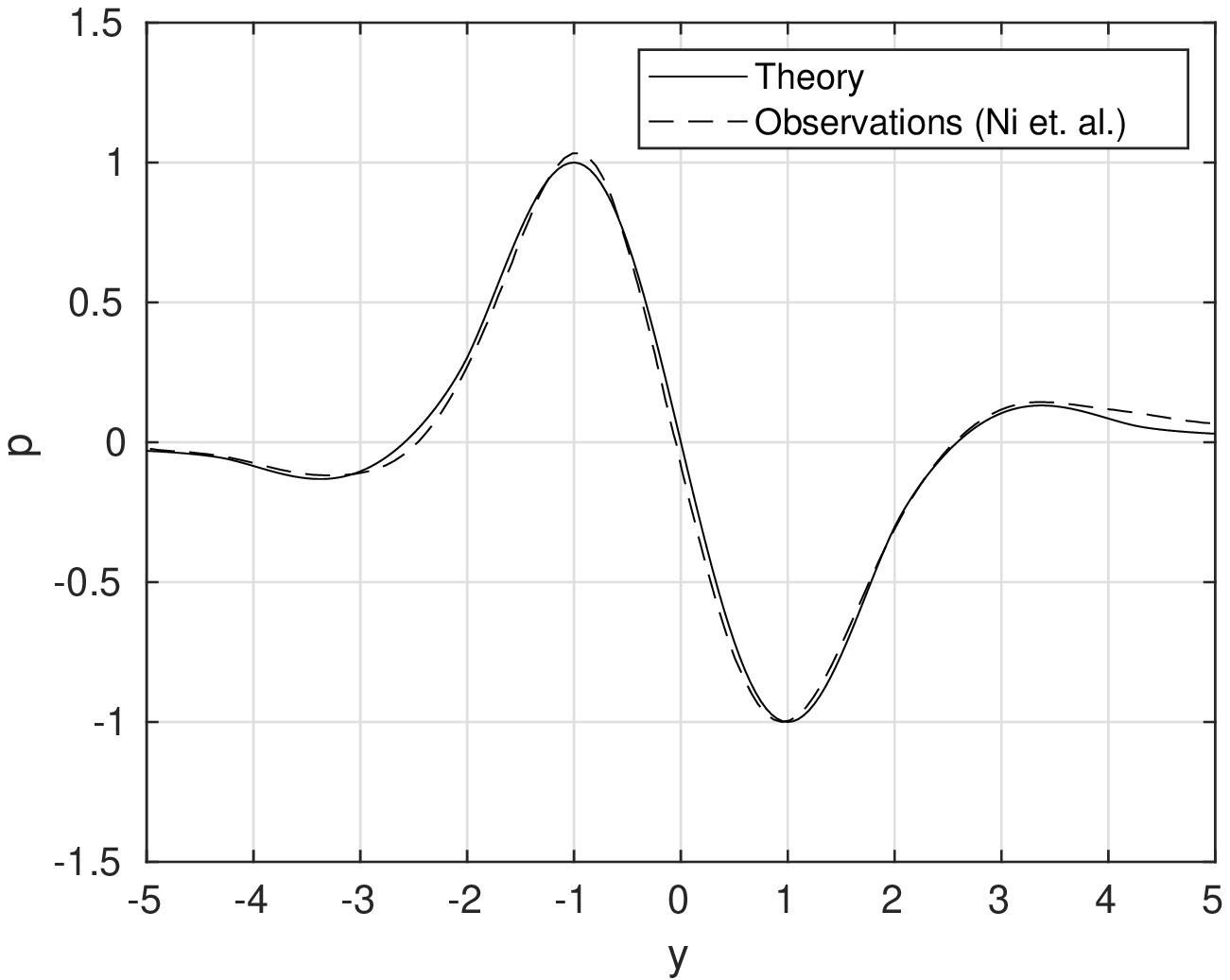}
     \hfill
     \includegraphics[width=0.4\textwidth]{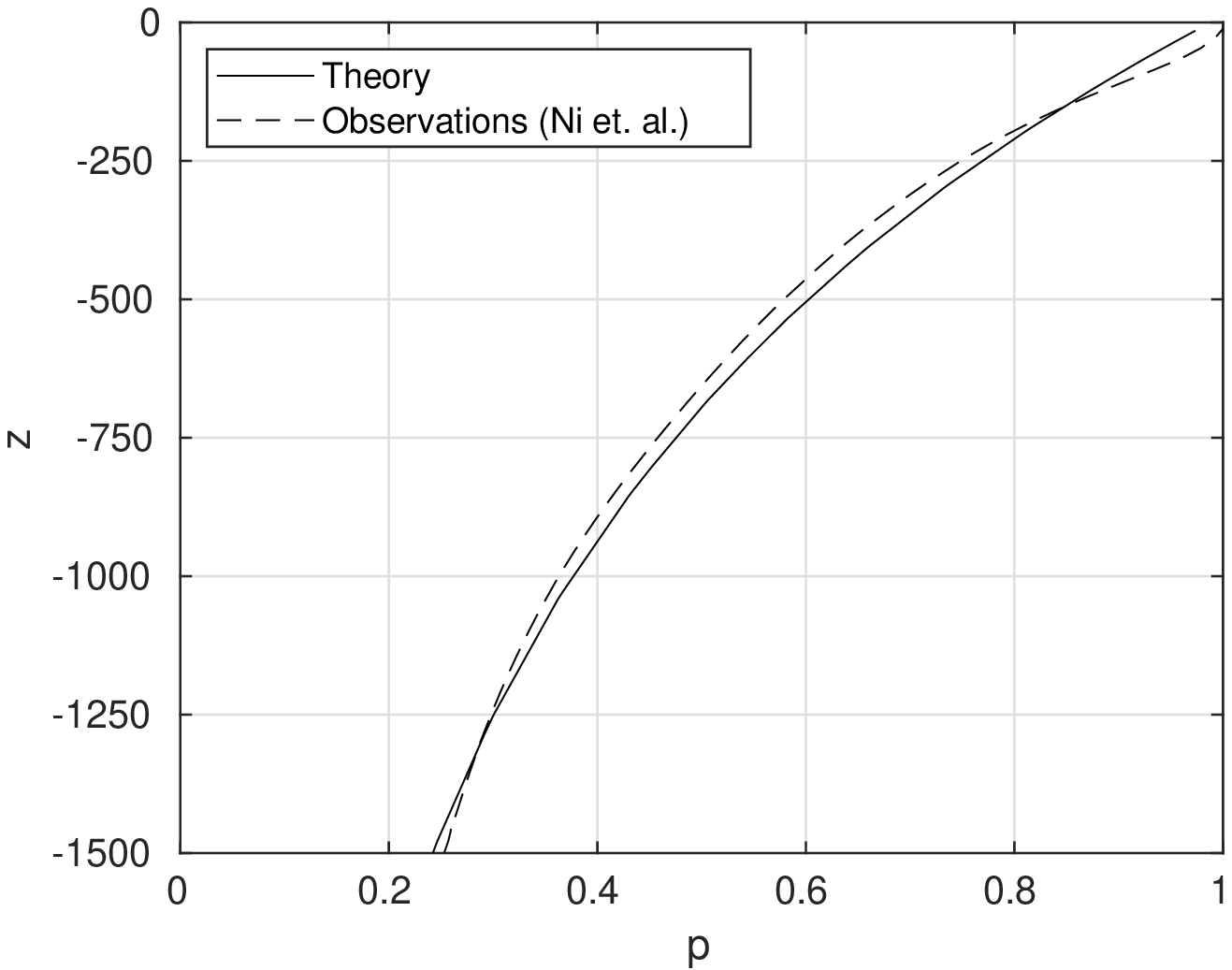}
     \hfill\phantom{.}
     \\
     \hspace*{0.25\textwidth}(a)\hspace{0.5\textwidth}(b)
\caption{Profiles of normalised pressure for the model (solid lines) and observed (dashed lines) composite dipoles. The model dipole consists of 80\% mode-one and 20\% mode-two modons. (a) horizontal, scaled on distance to the first vorticity maximum. (b) vertical, in metres.
\label{f:profiles}}
\end{figure}

It is possible that the composite dipole of NZWH might combine instantaneous observations of both mode-one and mode-two vortices. Figure \ref{f:profiles}(a) shows a comparison between the NZWH composite dipole, reproduced in figure \ref{f:Ni}(b,d) here, and a model composite formed from 80\% mode-one and 20\% mode-two of \S\ref{S:gov}. Following NZWH, we have rescaled both  modons so the first maximum occurs at $y = 1$ and the maximum pressure is unity. The model composite dipole is then determined by fitting an arbitrary linear combination of the two modes to the profiles in figure \ref{f:Ni} to give the optimum 80/20 composite used in figure \ref{f:profiles}(a) and subsequently.
Figure \ref{f:profiles}(b)  gives a comparison between the vertical structure of the model composite dipole and the reported vertical structure in NZWH, where the vertical decay scale for the model dipole is determined by fitting the model solution to the vertical profile from NZWH. The small discrepancies near the surface are probably due to the presence of a surface mixed layer in the observations.
\begin{figure}
\hfill
     \includegraphics[width=0.4\textwidth]{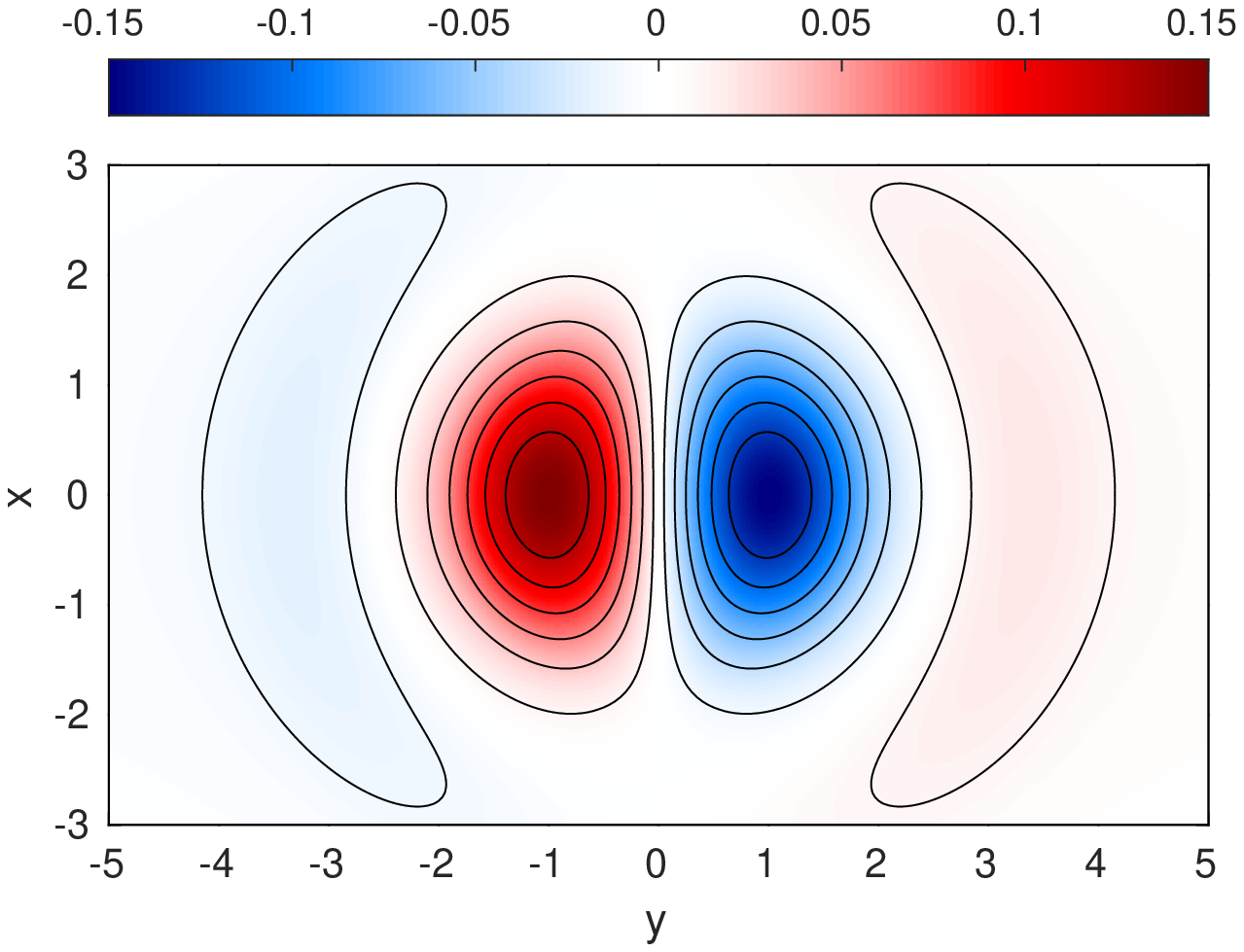}
     \hfill
     \includegraphics[width=0.4\textwidth]{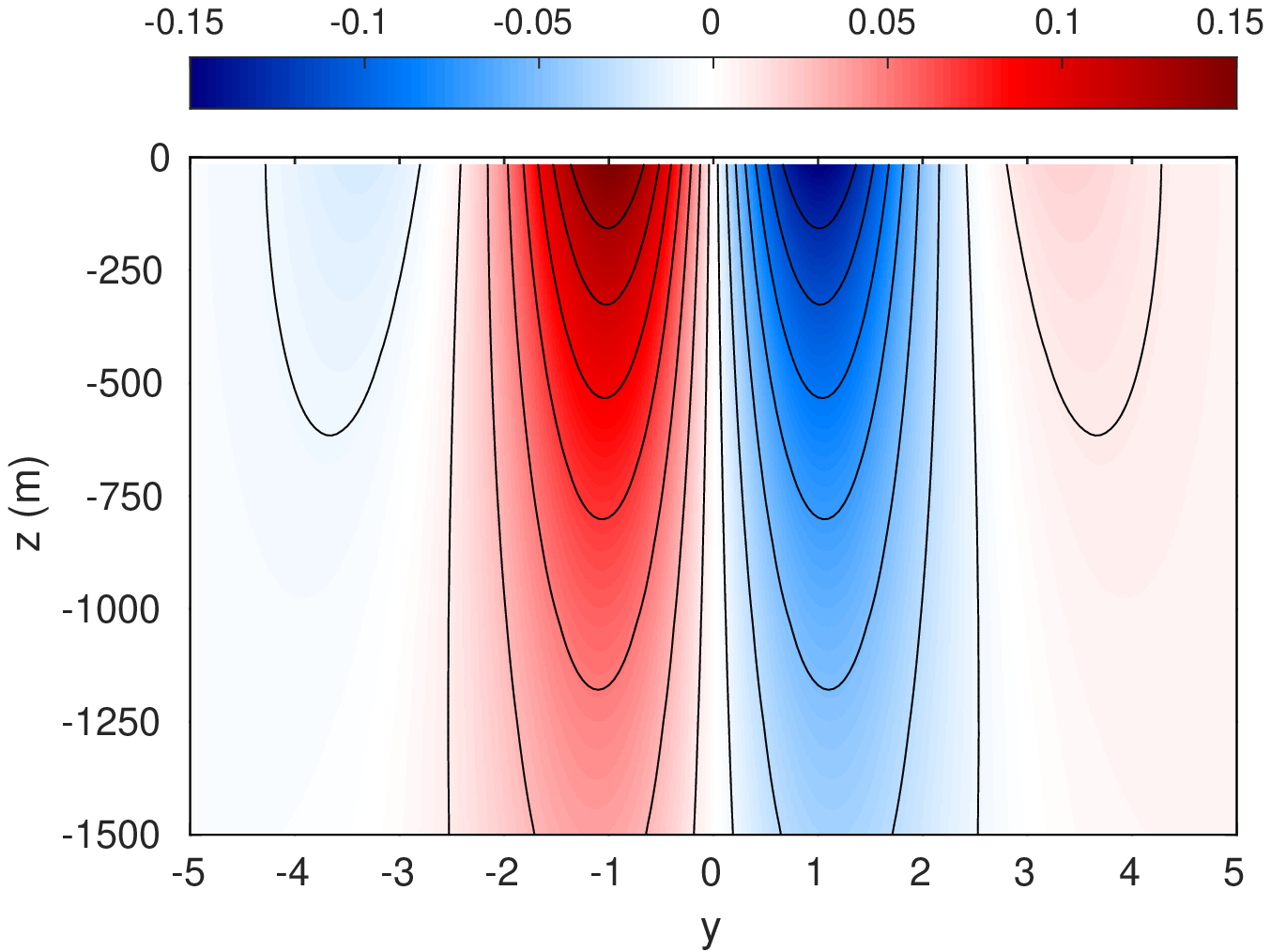}
     \hfill\phantom{.}
     \\
     \hspace*{0.25\textwidth}(a)\hspace{0.5\textwidth}(b)\\
\caption{Profiles of normalised pressure for the model composite dipole. (a) The surface pressure scaled on distance to the first vorticity maximum. (b) A vertical section along $x=0$ of the pressure with depth in metres.
\label{f:composite}}
\end{figure}

Figure \ref{f:composite} shows the full surface and vertical structure of the model composite.  Results have been scaled to match the maximum values and the vertical decay scale of NZWH reproduced in figure \ref{f:Ni}(a,c) here.
\begin{figure}
\hfill
     \includegraphics[width=0.45\textwidth]{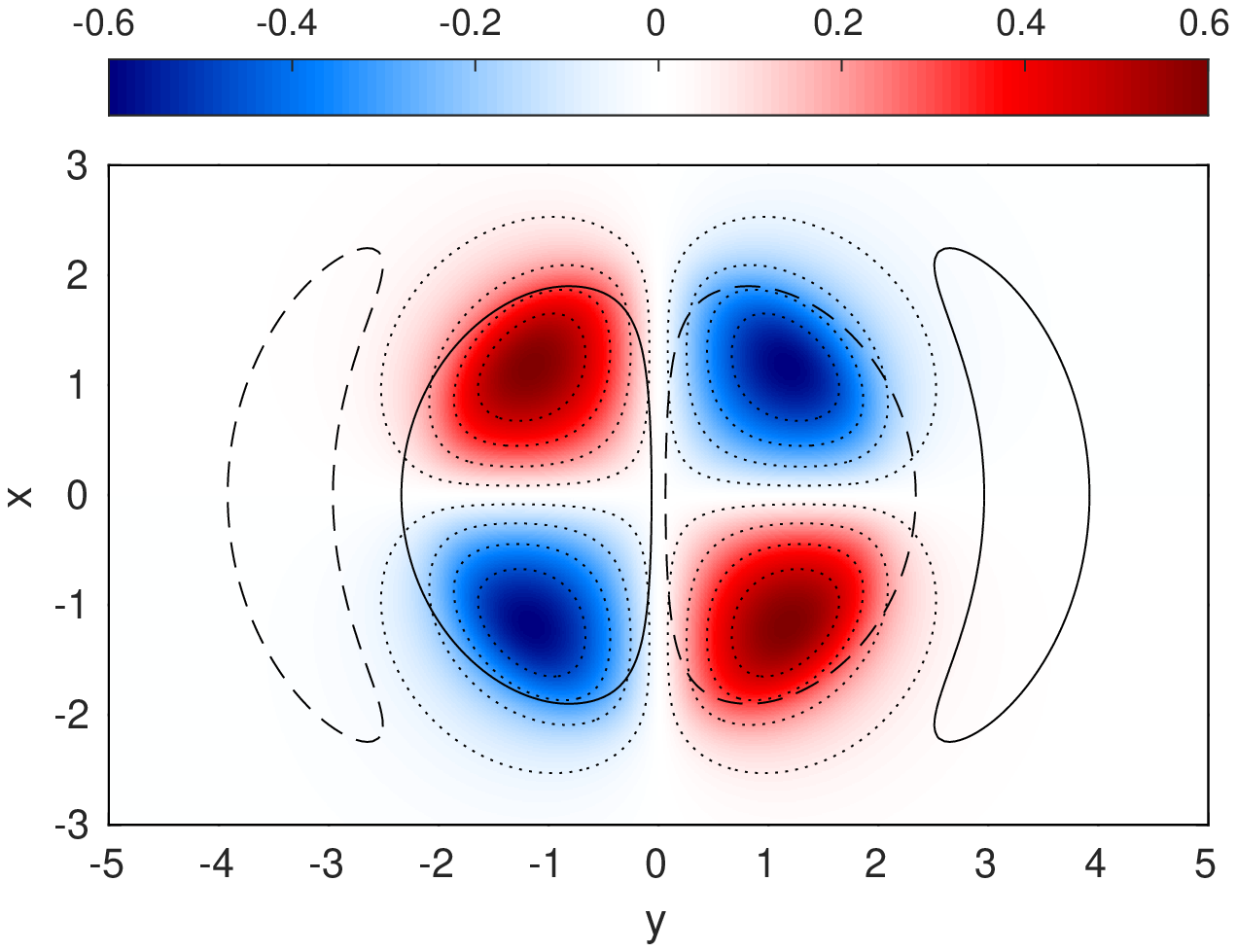}
     \hfill
     \includegraphics[width=0.45\textwidth]{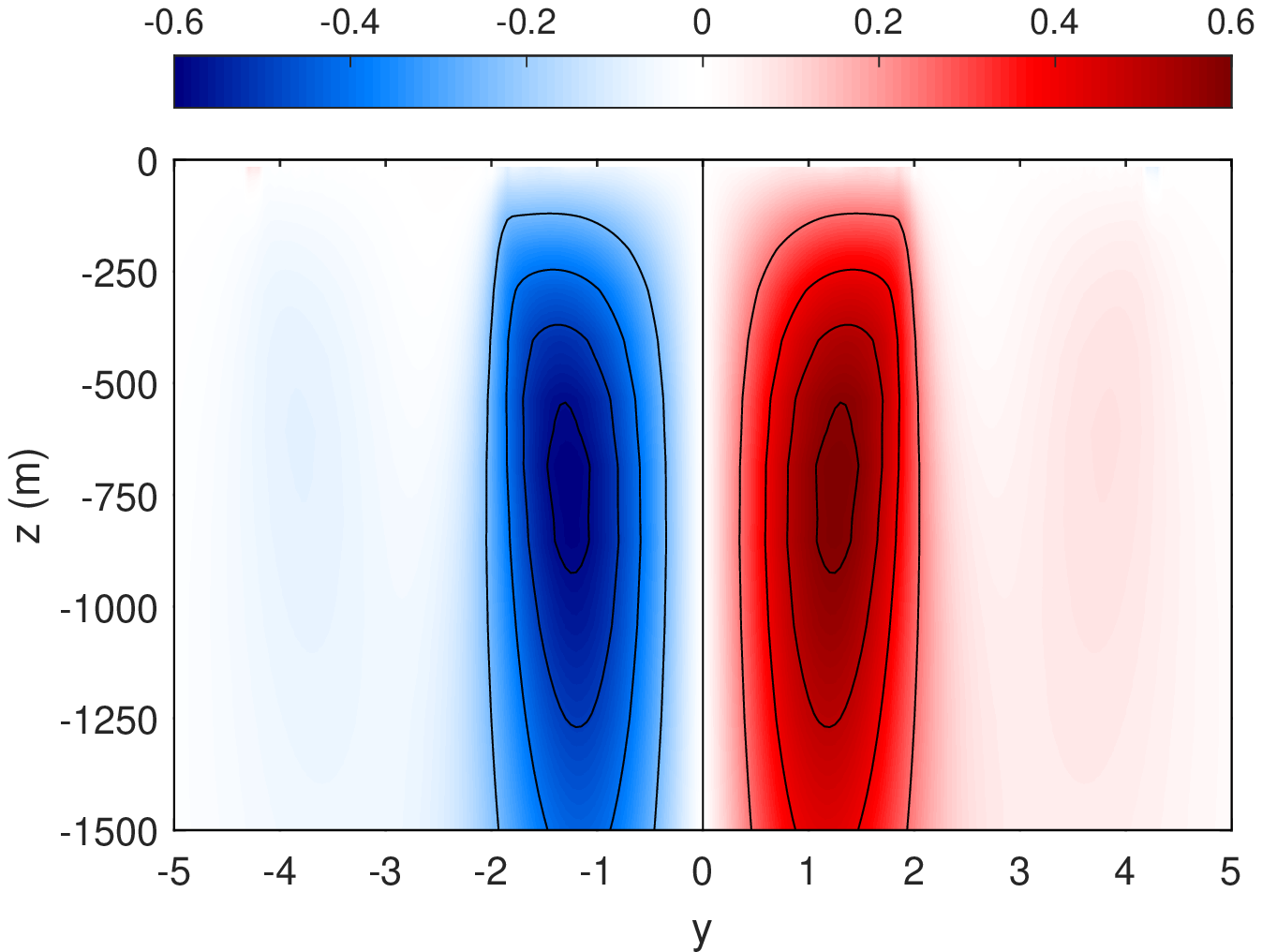}
     \hfill\phantom{.}
     \\
     \hspace*{0.25\textwidth}(a)\hspace{0.5\textwidth}(b)
\caption{The frontogenetic vertical velocity (m day$^{-1}$ ) inside the model composite dipole. (a) The horizontal pattern of $w$ at 680 m depth. (b) A vertical section along $x=-0.7$.}
\label{f:compvertvel}
\end{figure}
Figure \ref{f:compvertvel} shows the vertical velocity at a non-dimensional depth of $z = -0.185$ corresponding to a dimensional depth of $-680\,\textrm{m}$,  calculated from the density equation \eqref{density} which becomes here
\begin{equation}\label{mod_dens}
w = \partial(p_z,p+y),
\end{equation}
giving $w$ vanishing on the top surface ($z=0$) from \eqref{gov2_surf} by construction, and showing, as noted in \S\ref{S:gov}, that in this model the vertical velocity is determined by the balance between horizontal advection of disturbance density and vertical advection of background density. Figure \ref{f:compvertvel} agrees closely with NZWH figure 8 suggesting that this frontogenetic vertical velocity is well captured by surface quasigeostrophic dynamics. The maximum value of $w$ varies with the vortex speed and radius so we have taken a value here which corresponds to the value in NZWH figure 8.

\section{Discussion}

We have presented a fast and simple linear explicit method to solve the nonlinear problem for a surface geostrophic dipole posed by MS. The method gives higher order dipoles directly and we show that a composite model dipole formed from a combination of mode-one and mode-two dipoles fits well the composite dipole put forward from observations by NZWH. Vertical velocities in the model composite dipole arise from the balance between horizontal advection of disturbance density and vertical advection of background density and appear to fit well the observed frontogenetic velocities.\\

\noindent{\bf Funding\bf{.}} This work was funded by the UK Natural Environment Research Council under grant number NE/S009922/1.\\
\noindent{\bf Declaration of interests\bf{.}} The authors report no conflict of interest.\\
\noindent{\bf Acknowledgement\bf{.}} The authors are indebted to Prof. Christopher Hughes for comments on a previous draft of this work.

\appendix

\section{Finite depth and exponential stratification \label{S:expstrat}}

Observed profiles of buoyancy frequency vary significantly with depth.
In terms of the present variables, the profile proposed by \cite{GarrettM72} for $N$ below the mixed layer can be written
\begin{equation}
    N ( z ) = N_0 \exp (\mu z),
    \label{Garrett}
\end{equation}
where $N_0$ = $\hat{N}\exp ( - H / H_N )$ and $\mu = f L/N_0H_N$, for stratification scale height $H_N$. \cite{GarrettM72} give typical values as $\hat{N}= 3$ cph and $H_N = 1.3$km. For finite $B$ and density profile \eqref{Garrett}, equations \eqref{hankel}, \eqref{sbc2} become
\begin{equation} \label{hankelB}
 p(r,\phi,z) = \sin\phi\int_0^\infty \hat{p}(\xi)\J1(\xi r)k(\xi,z;\mu,B)\xi \;\rmd\xi,
\end{equation}
\begin{equation}\label{sbcB}
\int_0^\infty A(\xi)\J1(\xi r)\;\rmd\xi-K\int_0^\infty [k(\xi,0;\mu,B)/k_z(\xi,0;\mu,B)]A(\xi)\J1(\xi r)\;\rmd\xi = K r,  
\end{equation}
for kernel function \cite[][]{Johnson78b}
\begin{equation}
k(\xi,z;\mu,B) = e^{\mu  z} \frac{K_0(\xi e^{-\mu  B}/\mu ) I_1(\xi e^{\mu  z}/\mu)
 + I_0(\xi e^{-\mu  B}/\mu ) K_1(\xi e^{\mu  z}/\mu)}
{K_0(\xi e^{-\mu  B}/\mu ) I_1(\xi /\mu)
 + I_0(\xi e^{-\mu  B}/\mu ) K_1(\xi/\mu)}.
\end{equation}
The integrals corresponding to \eqref{bmndef}-\eqref{cmdef} must now be evaluated numerically but the remainder of the analysis is unaltered and the economy of the expansion and solution remains.

For weakly stratified where $B\ll1$ with $z/B$ fixed $k(\xi,0;\mu,B)/k_z(\xi,0;\mu,B)\to\xi^{-2}$, the inverse Laplace operator. The exact solution of \eqref{hankelB}, \eqref{sbcB} and \eqref{outer} follows as the depth-independent Lamb-Chaplygin vortex, as demonstrated numerically by MS.

\section{Evaluation of the series \eqref{pzZ} \label{S:series}}
The recurrence relation for the radial polynomials gives the three term recurrence relation
\begin{equation}
    \RR_n(r) = \alpha_n(r)\RR_{n-1}(r) + \gamma_n\RR_{n-2}(r),
\end{equation}
where
\begin{equation}
\alpha_n(r)=\frac{4n^2-(8n^2-2)r^2}{(2n-1)(n+1)}, \qquad \gamma_n=-\frac{(2n+1)(n-1)}{(2n-1)(n+1)}.
\end{equation}
The Clenshaw algorithm \cite[][\S5.4.2]{PressTVF07} then gives 
\begin{align}
    &S_{N+2} =0, \qquad S_{N+1} =0, \\
    &S_k = \alpha_{k+1}(r)S_{k+1} + \gamma_{k+2} S_{k+2} +a_k, \qquad k=N,N-1,\dots,1, \\
    &\sum_{n=0}^N a_n \RR_n(r) = S_1(2r-3r^3)+(a_0-5S_2/9)r.
\end{align}
This gives the surface PV as a polynomial in $(x,y)$,
\begin{equation}
   p_z(x,y) = \sin\phi \sum_{n=0}^N a_n \RR_n(r) = y [S_1(2-3r^2)+(a_0-5S_2/9)]. 
\end{equation}

\bibliographystyle{jfm.bst}
\bibliography{SQG.bib}

\begin{thebibliography}{14}
\expandafter\ifx\csname natexlab\endcsname\relax\def\natexlab#1{#1}\fi
\def\au#1{#1} \def\ed#1{#1} \def\yr#1{#1}\def\at#1{#1}\def\jt#1{\textit{#1}}
  \def\bt#1{#1}\def\bvol#1{\textbf{#1}} \def\vol#1{#1} \def\pg#1{#1}
  \def\publ#1{#1}\def\arxiv#1{#1}\def\org#1{#1}\def\st#1{\textit{#1}}

\bibitem[Born \& Wolf(2019)]{BornW19}
{\sc \au{Born, M.} \& \au{Wolf, E.}} \yr{2019} {\em Principles of Optics\/},
  7th edn.  \publ{CUP}.

\bibitem[Burns {\em et~al.\/}(2020)Burns, Vasil, Oishi, Lecoanet \&
  Brown]{BurnsVOLB20}
{\sc \au{Burns, K.~J.}, \au{Vasil, G.~M.}, \au{Oishi, J.~S.}, \au{Lecoanet, D.}
  \& \au{Brown, B.~P.}} \yr{2020}  \at{Dedalus: {A} flexible framework for
  numerical simulations with spectral methods}.  \jt{Phys. Rev. Res.}
  \bvol{2},  \pg{023068}.

\bibitem[Couder \& Basdevant(1986)]{Couder86}
{\sc \au{Couder, Y.} \& \au{Basdevant, C.}} \yr{1986}  \at{Experimental and
  numerical study of vortex couples in two-dimensional flows}.  \jt{J. Fluid
  Mech.}  \bvol{173},  \pg{225--251}.

\bibitem[Garrett \& Munk(1972)]{GarrettM72}
{\sc \au{Garrett, C.} \& \au{Munk, W.}} \yr{1972}  \at{Space-time scales of
  internal waves}.  \jt{Geophys. Fluid Dyn.}  \bvol{3},  \pg{225--264}.

\bibitem[Held {\em et~al.\/}(1995)Held, Pierrehumbert, Garner \&
  Swanson]{HeldPGS95}
{\sc \au{Held, I.~M.}, \au{Pierrehumbert, R.~T.}, \au{Garner, S.~T.} \&
  \au{Swanson, K.~L.}} \yr{1995}  \at{Surface quasi-geostrophic dynamics}.
  \jt{J. Fluid Mech.}  \bvol{282},  \pg{1--20}.

\bibitem[Hocking {\em et~al.\/}(1979)Hocking, Moore \& Walton]{HockingMW79}
{\sc \au{Hocking, L.~M.}, \au{Moore, D.~W.} \& \au{Walton, I.~C.}} \yr{1979}
  \at{Drag on a sphere moving axially in a long rotating container}.  \jt{J.
  Fluid Mech.}  \bvol{90},  \pg{781--793}.

\bibitem[Johnson(1978)]{Johnson78b}
{\sc \au{Johnson, E.~R.}} \yr{1978}  \at{Topographically bound vortices}.
  \jt{Geophys Astrophys Fluid Dyn}  \bvol{11},  \pg{61-- 71}.

\bibitem[Lahaye {\em et~al.\/}(2020)Lahaye, Zeitlin \& Dubos]{LahayeZD20}
{\sc \au{Lahaye, N.}, \au{Zeitlin, V.} \& \au{Dubos, T.}} \yr{2020}
  \at{Coherent dipoles in a mixed layer with variable buoyancy: Theory compared
  to observations}.  \jt{Ocean Modelling}  \bvol{153},  \pg{101673}.

\bibitem[Meleshko \& van Heijst(1994)]{MeleshkoH94}
{\sc \au{Meleshko, V.~V.} \& \au{van Heijst, G. J.~F.}} \yr{1994}  \at{On
  {C}haplygin's investigations of two-dimensional vortex structures in an
  inviscid fluid}.  \jt{J. Fluid Mech.}  \bvol{272},  \pg{157--182}.

\bibitem[Muraki \& Snyder(2007)]{MurakiS07}
{\sc \au{Muraki, D.~J.} \& \au{Snyder, C.}} \yr{2007}  \at{Vortex dipoles for
  surface quasigeostrophic models}.  \jt{J. Atmospheric Sci.}  \bvol{64},
  \pg{2961--2967}.

\bibitem[Ni {\em et~al.\/}(2020)Ni, Zhai, Wang \& Hughes]{NiZWH20}
{\sc \au{Ni, Q.}, \au{Zhai, X.}, \au{Wang, G.} \& \au{Hughes, C.~W.}} \yr{2020}
   \at{Widespread mesoscale dipoles in the global ocean}.  \jt{J. Geophys. Res.
  Oceans}  \bvol{125},  \pg{e2020JC016479}.

\bibitem[Press {\em et~al.\/}(2007)Press, Teukolsky, Vetterling \&
  Flannery]{PressTVF07}
{\sc \au{Press, W~H}, \au{Teukolsky, S~A}, \au{Vetterling, W~T} \&
  \au{Flannery, B~P}} \yr{2007} {\em Numerical Recipes: The Art of Scientific
  Computing\/}.  \publ{CUP}.

\bibitem[Tranter(1971)]{Tranter71}
{\sc \au{Tranter, C.~J.}} \yr{1971} {\em Integral Transforms in Mathematical
  Physics\/}.  \publ{Methuen}.

\bibitem[Warneford \& Dellar(2013)]{WarnefordD13}
{\sc \au{Warneford, E.~S.} \& \au{Dellar, P.~J.}} \yr{2013}  \at{The
  quasi-geostrophic theory of the thermal shallow water equations}.  \jt{J.
  Fluid Mech.}  \bvol{723},  \pg{374–403}.

\end{thebibliography}

\end{document}